\documentclass[12pt]{article}

\usepackage{axodraw}

\begin{document}
\begin{titlepage}
\begin{flushright}
UFIFT-QG-04-03 \\ UFIFT-HEP-05-07 \\ hep-ph/0505156
\end{flushright}
\vspace{.4cm}
\begin{center}
\textbf{Yukawa Scalar Self-Mass on a Conformally Flat Background}
\end{center}
\begin{center}
L. D. Duffy$^{\dagger}$ and R. P. Woodard$^{\ddagger}$
\end{center}
\begin{center}
\textit{Department of Physics \\ University of Florida \\
Gainesville, FL 32611 USA}
\end{center}

\begin{center}
ABSTRACT
\end{center}

We compute the one loop self-mass-squared of a massless, minimally 
coupled scalar which is Yukawa-coupled to a massless Dirac fermion 
in a general conformally flat background. Dimensional regularization 
is employed and a fully renormalized result is obtained. For the 
special case of a locally de Sitter background our result is 
manifestly de Sitter invariant. By solving the effective field
equations we show that the scalar mode functions acquire no 
significant one loop corrections. In particular, the phenomenon of 
super-adiabatic amplification is not affected. One consequence is 
that the scalar-catalyzed production of fermions during inflation 
should not be reduced by changes in the scalar sector before it has
time to go to completion.

\begin{flushleft}
\end{flushleft}
\vspace{.4cm}
\begin{flushleft}
$^{\dagger}$ e-mail: leanned@phys.ufl.edu \\
$^{\ddagger}$ e-mail: woodard@phys.ufl.edu
\end{flushleft}
\end{titlepage}

\section{Introduction}

Essentially all quantum field theoretic effects can be understood
through classical interactions of the virtual particles whose existence
is required by the uncertainty principle. In general one expects quantum
field theoretic effects to become stronger the longer virtual particles 
live and the more probable it is for them to emerge from the vacuum.
For example, vacuum polarization arises due to the polarization of
charged virtual particles in an external electric field. The largest
effect derives from electron-positron pairs because they are the lightest
charged particles and therefore live the longest. One can even understand
the running of the electromagnetic force from the incomplete polarization 
of the longest wavelength virtual pairs in the field provided by two very 
close sources.

The expansion of spacetime affects quantum field theory by lengthening the
time virtual particles can exist, and sometimes by altering the probability 
with which they emerge from the vacuum. The first effect can be understood
from the energy-time uncertainty principle. In co-moving coordinates
the geometry of a homogeneous, isotropic and spatially flat universe is,
\begin{equation}
ds^2 = -dt^2 + a^2(t) d\vec{x} \cdot d\vec{x} \; . 
\end{equation}
Although $t$ measures physical time, physical distance is $\Delta x$
multiplied by the scale factor $a(t)$. Because spatial translation 
invariance is still a good symmetry one can label particles by their wave 
vectors $\vec{k}$ as in flat space. However, the physical momentum of such
a particle is $\vec{k}/a(t)$, and one can think of the corresponding 
energy as,
\begin{equation}
E(\vec{k},t) = \sqrt{m^2 + \Vert \vec{k} \Vert^2/a^2(t)} \; .
\end{equation}
The spontaneous emergence of a pair with wave numbers $\pm \vec{k}$ at 
time $t$ will not lead to a detectable violation of energy conservation
provided the pair persists no longer than a time $\Delta t$ defined by
the equation,
\begin{equation}
\int_t^{t+{\Delta t}} \!\!\!\!\!\! dt' \, 2 E(\vec{k},t') = 1 \; .
\end{equation}
Hence we conclude that the expansion of spacetime always increases the
time which a virtual pair can persist.

Just as in flat space the persistence time $\Delta t$ is longest for
massless particles. In an expanding geometry this means $m \ll H(t)$,
where the Hubble and deceleration parameters are,
\begin{equation}
H(t) \equiv \frac{\dot{a}}{a} \qquad , \qquad q(t) \equiv - 
\frac{a \ddot{a}}{\dot{a}^2} = -1 - \frac{\dot{H}}{H^2} \; .
\end{equation}
In this case the equation for persistence time takes the form,
\begin{equation}
m = 0 \qquad \Longrightarrow \qquad 2 \Vert \vec{k} \Vert 
\int_t^{t+{\Delta t}} \frac{dt'}{a(t')} = 1 \; . \label{m=0}
\end{equation}
The integral in (\ref{m=0}) is just the conformal time interval
$\Delta \eta$, defined by the change of variables $d\eta = dt/a(t)$. The
dependence of $\Delta \eta$ upon $\Delta t$ is controlled by sign of $q(t)$.
For $q(t) > 0$ (decelerating expansion) $\Delta \eta$ grows without bound; 
while for $q(t) < 0$ (inflation) it approaches a finite constant. Hence we 
conclude that any sufficiently long wavelength virtual particles which are 
produced during inflation can persist forever!

Whether or not effectively massless particles actually engender stronger
quantum effects during inflation depends upon the probability with which
they emerge from the vacuum. Almost all massless particles possess a 
symmetry known as conformal invariance which means that physical processes
are the same as in flat space when expressed in conformal coordinates,
\begin{equation}
d\eta = \frac{dt}{a(t)} \qquad \Longrightarrow \qquad ds^2 = a^2 \Bigl(
-d\eta^2 + d\vec{x} \cdot d\vec{x}\Bigr) \; . 
\end{equation}
Hence the number of virtual particles which emerge from the vacuum per
conformal time is the same as the constant flat space rate we might
call $\Gamma$. It follows that the rate per physical time falls off,
\begin{equation}
\frac{dN}{dt} = \frac{d\eta}{dt} \, \frac{dN}{d\eta} = \frac{\Gamma}{a(t)} \; .
\end{equation}
Hence we conclude that while any sufficiently long wavelength, massless
virtual particles which happen to emerge from the vacuum can persist
forever during inflation, very few conformally invariant particles will
emerge. 

The two massless particles which are not conformally invariant are minimally
coupled scalars and gravitons. For these particles it turns out that the rate
of emergence per unit physical time is unsuppressed, so we expect quantum field
theoretic effects from them to be stronger than in flat space. In fact it is
quantum fluctuations of these fields which are responsible for the scalar
\cite{Slava} and tensor \cite{Alexei} perturbations predicted by inflation
\cite{MFB,LL}. These are tree order effects. At two loop order it can be shown
that a massless, minimally coupled scalar with a quartic self-interaction
experiences violations of the weak energy condition on cosmological scales
\cite{OW1,OW2} and that quantum gravitational back-reaction slows inflation
\cite{TW}.

Conformal invariance in the free theory need not rule out significant
quantum corrections if the conformally invariant particle couples to one
which is not conformally invariant. Two examples of this have been studied
recently. In the first, electrodynamics --- which is conformally invariant
in $D=4$ spacetime dimensions --- is coupled to a charged, massless,
minimally coupled scalar. The one loop vacuum polarization \cite{PTW1,PTW2}
induced by the latter causes super-horizon photons to behave, in some
ways, as though they had nonzero mass \cite{PW1}. This engenders no
physical photon creation during inflation but leads instead to a vast
enhancement of the 0-point energy of super-horizon photons which
may seed cosmic magnetic fields after the end of inflation \cite{DDPT,DPTD}.

\begin{center} 
\begin{picture}(300,100)(0,0)
\ArrowLine(70,50)(20,50)
\Text(25,55)[b]{$i$}
\Vertex(70,50){2}
\Text(70,43)[t]{$x$}
\ArrowLine(130,50)(70,50)
\Vertex(130,50){2}
\Text(130,46)[t]{$x'$}
\ArrowLine(180,50)(130,50)
\Text(175,55)[b]{$j$}
\DashCArc(100,50)(30,0,180){2}
\Text(200,46)[b]{\Large $+$}
\ArrowLine(260,50)(220,50)
\Text(225,55)[b]{$i$}
\Vertex(260,50){2}
\Text(261,45)[b]{\Large $\times$}
\Text(260,56)[b]{$x$}
\ArrowLine(300,50)(260,50)
\Text(295,55)[b]{$j$}
\end{picture} 
\\ {\rm Fig.~1: One loop contributions to $\Bigl[{}_i \Sigma_j\Bigr](x;x')$.} 
\end{center}

The other example consists of a massless fermion which is Yukawa-coupled
to a massless, minimally coupled scalar \cite{PW2}. Massless fermions are
conformally invariant (in any dimension) so they are not much produced by
themselves. However, the fermion self-energy (Fig.~1), and the effect it
has on the quantum-corrected fermion field equations, show that the massless, 
minimally coupled scalar catalyzes the emission of scalar-fermion-anti-fermion 
from the vacuum. In this paper we study what effect the fermion has on the 
scalar through the one loop self-mass-squared (Fig.~2). This can hardly alter 
the non-conformal coupling but it {\it might} induce a nonzero scalar mass. 
If this mass became large enough sufficiently quickly it could cut off the 
scalar-induced fermion creation. We will show that this does not happen. 
The one loop scalar self-mass-squared can be renormalized so that there is 
no significant change to the scalar mode functions, and higher loop 
corrections cannot become significant soon enough.

In the next section we use the Yukawa Lagrangian to derive the relevant
Feynman rules for an arbitrary scale factor $a(t)$. In section 3 we
compute the renormalized scalar self-mass-squared at one loop order.
Although our result is valid for any $a(t)$ we show that it reduces
to a manifestly de Sitter invariant form for the locally de Sitter
case of $a(t) = e^{H t}$. In section 4 we use the self-mass-squared
to study one loop corrections to the scalar mode functions. The 
consequences of this result are discussed in section 5.

\begin{center} 
\begin{picture}(300,100)(0,0)
\DashLine(20,50)(70,50){2}
\Vertex(70,50){2}
\Text(75,50)[l]{$x$}
\DashLine(130,50)(180,50){2}
\Vertex(130,50){2}
\Text(125,51)[r]{$x'$}
\ArrowArc(100,50)(30,0,180)
\ArrowArc(100,50)(30,180,360)
\Text(200,46)[b]{\Large $+$}
\DashLine(220,50)(300,50){2}
\Vertex(260,50){2}
\Text(261,45)[b]{\Large $\times$}
\Text(260,56)[b]{$x$}
\end{picture} 
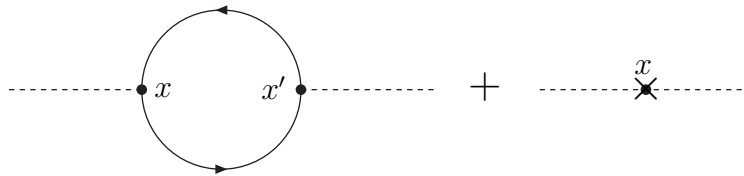
\\ {\rm Fig.~2: One loop contributions to $M^2(x;x')$.} 
\end{center}

\section{Feynman Rules}

We begin by reviewing the conventions appropriate to Dirac fields in a
nontrivial geometry. In order to facilitate dimensional regularization
we make no assumption about the spacetime dimension $D$. The gamma matrices
$\gamma^{b}_{ij}$ ($b=0,1,\dots, D\!-\!1$) anti-commute in the usual way,
$\{\gamma^b,\gamma^c\} = -2 \eta^{bc} I$. One interpolates between local
Lorentz indices ($b,c,d,\dots$) and vector indices (lower case Greek letters)
with the vierbein field, $e_{\mu b}(x)$. The metric is obtained by contracting
two vierbeins with the Minkowski metric, $g_{\mu\nu}(x) = e_{\mu b}(x)
e_{\nu c}(x) \eta^{bc}$. The vierbein's vector index is raised and lowered by
the metric ($e^{\mu}_{~b} = g^{\mu\nu} e_{\nu b}$) while the local Lorentz
index is raised and lowered with the Minkowski metric ($e_{\mu}^{~b} =
\eta^{bc} e_{\mu c}$). The spin connection and the Lorentz representation
matrices are,
\begin{equation}
A_{\mu bc} \equiv e^{\nu}_{~b} \Bigl(e_{\nu c , \mu} - \Gamma^{\rho}_{~\mu \nu}
e_{\rho c}\Bigr) \qquad , \qquad J^{bc} \equiv \frac{i}4 [\gamma^b,\gamma^c]
\; .
\end{equation}

Let $\phi(x)$ represent a real scalar field and let $\psi_i(x)$ stand for a
Dirac field. In a general background metric the Lagrangian we wish to study
would be,
\begin{eqnarray}
\lefteqn{{\cal L} = -\frac12 \partial_{\mu} \phi \partial_{\nu} \phi
g^{\mu\nu} \sqrt{-g} + \overline{\psi} e^{\mu}_{~b} \gamma^b \Bigl(i 
\partial_{\mu} - \frac12 A_{\mu cd} J^{cd} \Bigr) \psi \sqrt{-g} } \nonumber \\
& & \hspace{5cm} - \frac12 \xi_0 \phi^2 R \sqrt{-g} - f_0 \phi \overline{\psi}
\psi \sqrt{-g} \; ,
\end{eqnarray}
where $\overline{\psi} \equiv \psi^{\dagger} \gamma^0$ is the usual Dirac
adjoint, $\xi_0$ is the bare conformal coupling and $f_0$ is the bare Yukawa
coupling constant. Note that we do not require mass counterterms because
mass is multiplicatively renormalized in dimensional regularization.

The geometry of interest is the very special form associated with the 
homogeneous and isotropic element. By defining a new time coordinate 
$d\eta \equiv dt/a(t)$, the metric of this geometry can be 
made conformal to the Minkowski metric,
\begin{equation}
ds^2 = a^2 \Bigl(-d\eta^2 + d\vec{x} \cdot d\vec{x} \Bigr) \; . \label{conf}
\end{equation}
A convenient choice for the associated vierbein is, $e_{\mu b} = a
\eta_{\mu b}$. With these simplifications the spin connection assumes the
form,
\begin{equation}
e_{\mu b} = a \eta_{\mu b} \Longrightarrow A_{\mu cd} = \Bigl(\eta_{\mu c}
\partial_d - \eta_{\mu d} \partial_c \Bigr) \ln(a) \; .
\end{equation}
And our Lagrangian reduces to,
\begin{eqnarray}
\lefteqn{{\cal L} \rightarrow -\frac12 a^{D\!-\!2} \partial_{\mu} \phi
\partial_{\nu} \phi \eta^{\mu\nu} \!+\! \Bigl(a^{\frac{D-1}2} 
\overline{\psi}\Bigr) i \gamma^{\mu} \partial_{\mu} \Bigl(a^{\frac{D-1}2} 
\psi \Bigr) } \nonumber \\
& & + \frac12 \xi_0 (D\!-\!1) \Bigl(\!2 a_{,\mu\nu} a^{D\!-\!3} \!+\! (D-4) 
a_{,\mu} a_{,\nu} a^{D\!-\!4}\!\Bigr) \eta^{\mu\nu} \phi^2 - f_0 a^D \phi
\overline{\psi} \psi . \quad \label{L}
\end{eqnarray}

Renormalization is facilitated by introducing the renormalized fields,
\begin{equation}
\phi \equiv \sqrt{Z} \phi_R \qquad {\rm and} \qquad \psi \equiv \sqrt{Z_2}
\psi_R \; .
\end{equation}
This brings the Lagrangian to the form,
\begin{eqnarray}
\lefteqn{{\cal L} \rightarrow -\frac12 Z a^{D-2} \partial_{\mu} \phi_R
\partial^{\mu} \phi_R + Z_2 \Bigl( a^{\frac{D-1}2} 
\overline{\psi}_R \Bigr) i \hspace{-.15cm} \not \hspace{-.1cm} \partial 
\Bigl( a^{\frac{D-1}2} \psi_R \Bigr) } \nonumber \\
& & \hspace{-.5cm} + \frac12 Z \xi_0 (D\!-\!1) a^{D-2} \Bigl(2 
\frac{\partial^2 a}{a} \!+\! (D\!-\!4) \frac{\partial_{\mu} a}{a} 
\frac{\partial^{\mu} a}{a} \Bigr) \phi_R^2 - \sqrt{Z} Z_2 f_0 a^D \phi_R 
\overline{\psi}_R \psi_R . \qquad
\end{eqnarray}
Note the Dirac slash notation ($\not \hspace{-.1cm} \partial \equiv 
\gamma^{\mu} \partial_{\mu}$) and the convention --- used henceforth --- 
that indices are raised and lowered with the Lorentz metric (e.g., 
$\partial^{\mu} \equiv \eta^{\mu\nu} \partial_{\nu}$). Note also that 
$\partial^2 \equiv \partial_{\mu} \partial^{\mu}$. We now define the 
counterterms,
\begin{eqnarray}
Z \equiv 1 + \delta Z & , & Z_2 \equiv 1 + \delta Z_2 \; , \\
\sqrt{Z} Z_2 f_0 \equiv f + \delta f & , & Z \xi_0 \equiv 0 + \delta \xi \; .
\end{eqnarray}
Note that the conformal coupling enters only as a counterterm because we
want the scalar to be minimally coupled. We can now express the Lagrangian
in terms of primitive interactions and counterterms,
\begin{eqnarray}
\lefteqn{{\cal L} \rightarrow -\frac12 a^{D-2} \partial_{\mu} \phi_R
\partial^{\mu} \phi_R + \Bigl( a^{\frac{D-1}2} 
\overline{\psi}_R \Bigr) i \hspace{-.15cm} \not \hspace{-.1cm} \partial \Bigl( 
a^{\frac{D-1}2} \psi_R \Bigr) - f a^D \phi_R \overline{\psi}_R \psi_R } 
\nonumber \\
& & -\frac12 \delta Z a^{D-2} \partial_{\mu} \phi_R \partial^{\mu} \phi_R 
+ \frac12 \delta \xi (D\!-\!1) a^{D-2} \Bigl(2 \frac{\partial^2 a}{a} \!+\! 
(D\!-\!4) \frac{\partial_{\mu} a}{a} \frac{\partial^{\mu} a}{a} \Bigr) 
\phi_R^2 \qquad \nonumber \\
& & \hspace{3cm} + \delta Z_2 \Bigl( a^{\frac{D-1}2} \overline{\psi}_R \Bigr) i 
\hspace{-.15cm} \not \hspace{-.1cm} \partial \Bigl( a^{\frac{D-1}2} \psi_R 
\Bigr) - \delta f a^D \phi_R \overline{\psi}_R \psi_R . \label{cterms}
\end{eqnarray}
We shall need the counterterms on the second line; we will not need those
on the third line.

The diagrams of Fig.~2 do not require the scalar propagator. The fermion
propagator can be determined by noting from the second term of (\ref{cterms}) 
that the combination $a^{\frac{D-1}2} \psi_R$ behaves like a massless Dirac
field in flat space. It follows that the Feynman propagator of $\psi$ is just 
a conformal rescaling of the flat space result,
\begin{eqnarray}
i\Bigl[ {}_iS_j\Bigr](x;x') & = & (a a')^{\frac{1-D}2} \gamma^{\mu}_{ij} \, i
\partial_{\mu} \left\{ \frac{\Gamma(\frac{D}2 -1)}{4 \pi^\frac{D}2}
\Bigl[{\Delta x}^2(x;x') \Bigr]^{1-\frac{D}2} \right\} \; , \\
& = & \frac{\Gamma(\frac{D}2)}{2 \pi^\frac{D}2} (a a')^{\frac{1-D}2}
\frac{-i \gamma^{\mu}_{ij} {\Delta x}_{\mu}}{[{\Delta x}^2(x;x')]^{\frac{D}2}}
\; . \label{Sij}
\end{eqnarray}
Here ${\Delta x}_{\mu} \equiv \eta_{\mu\nu} (x^{\nu} \!-\! x^{\prime \nu})$
and the conformal coordinate interval is,
\begin{equation}
{\Delta x}^2(x;x') \equiv \Vert \vec{x} \!-\! \vec{x}' \Vert^2 - (\vert
\eta \!-\! \eta'\vert - i \delta)^2 \; . \label{Dx^2}
\end{equation}
Note that we label the spacetime position with the $D$-vector $x^{\mu}~
=~(\eta,\vec{x})$.
The split index notation in $i [{}_iS_j](x;x')$ indicates that the first index
($i$) transforms according to the local Lorentz group at the first coordinate
argument ($x^{\mu}$) whereas the second index ($j$) transforms at the second
argument ($x^{\prime \mu}$).

The interaction vertex derives from the $-f a^D \phi_R \overline{\psi}_R 
\psi_R$ term of (\ref{cterms}),
\begin{equation}
-i f a^D \delta_{ij} \; . \label{verts}
\end{equation}
We also require the scalar field strength renormalization and conformal
counterterms,
\begin{equation}
i \delta Z \partial_{\mu} \Bigl(a^{D\!-\!2} \partial^{\mu}\delta^D (x-x')
\Bigr) \; , \label{eq:delZcounterterm}
\end{equation}
and,
\begin{equation}
i \delta \xi (D\!-\!1) a^{D-2} \Bigl( 2 \frac{\partial^2 a}{a} 
+ (D-4) \frac{\partial_{\mu} a}{a} \frac{\partial^{\mu} a}{a} \Bigr) 
\delta^D(x-x') \; . \label{eq:delXicounterterm}
\end{equation}
Let us note, for future reference, that the choice,
\begin{equation}
\delta \xi = \frac14 \Bigl(\frac{D\!-\!2}{D\!-\!1}\Bigr) \delta Z \; ,
\label{special}
\end{equation}
makes the two counterterms sum to a simple form,
\begin{eqnarray}
\lefteqn{i \delta Z \partial_{\mu} \Bigl(a^{D\!-\!2} \partial^{\mu}
\delta^D(x-x') \Bigr) + i \delta Z a^{D-2} \Bigl(\frac{D}2 \!-\! 1\Bigl)
\left\{ \frac{\partial^2 a}{a} \right. } \nonumber \\
& & \hspace{1cm} \left. + \Bigl(\frac{D}2 \!-\! 2\Bigr) \frac{\partial_{\mu} 
a}{a} \frac{\partial^{\mu} a}{a} \right\} \delta^D(x-x') = i \delta Z 
(aa')^{\frac{D}{2}\!-\!1} \partial^2 \delta^D(x-x') \; . \qquad \label{sum}
\end{eqnarray}

Because the ``in'' ($t\rightarrow -\infty$) vacuum is not equal to 
the ``out'' (t$\rightarrow +\infty$) vacuum in this background it is
desirable to compute true expectation values rather than in-out matrix
elements. This can be done covariantly using a simple extension of the
Feynman rules known as the Schwinger-Keldysh formalism 
\cite{Schwinger:1960qe,M,BM,K}. Briefly, the end of each line has a
polarity which can be ``$+$'' or ``$-$''. Vertices are either all $+$
or all $-$. A $+$ vertex is the familiar one from the Feynman rules 
whereas the $-$ vertex is its negative. Propagators can be 
$++$, $+-$, $-+$ or $--$. Each propagator can be obtained from the Feynman 
propagator by replacing the conformal coordinate interval, ${\Delta x
}^2(x;x')$, with the interval of appropriate polarization,
\begin{eqnarray}
{\Delta x}^{2}_{\scriptscriptstyle ++}(x;x') &\equiv& \Vert \vec{x} \!-\! 
\vec{x}' \Vert^2 - (\vert \eta \!-\! \eta'\vert - i \delta)^2 \; , 
\label{Dx++^2} \\
{\Delta x}^{2}_{\scriptscriptstyle +-}(x;x') &\equiv& \Vert \vec{x} \!-\! 
\vec{x}' \Vert^2 - (\eta \!-\! \eta' + i \delta)^2 \; , \label{Dx+-^2} \\
{\Delta x}^{2}_{\scriptscriptstyle -+}(x;x') &\equiv& \Vert \vec{x} \!-\! 
\vec{x}' \Vert^2 - (\eta \!-\! \eta' - i \delta)^2 \; , \label{Dx-+^2} \\
{\Delta x}^{2}_{\scriptscriptstyle --}(x;x') & \equiv & \Vert \vec{x} \!-\! 
\vec{x}' \Vert^2 - (\vert \eta \!-\! \eta' \vert + i \delta)^2 \; . 
\label{Dx--^2}
\end{eqnarray}

External lines can be either $+$ or $-$ in the Schwinger-Keldysh formalism.
Hence every N-point 1-particle-irreducible (1PI) function of the in-out
formalism gives rise to $2^N$ 1PI functions in the Schwinger-Keldysh 
formalism. For every field $\phi(x)$ of an in-out effective action, 
a Schwinger-Keldysh effective action must depend upon two fields
--- call them $\phi_+(x)$ and $\phi_-(x)$ --- in order to access the
appropriate 1PI function \cite{CSHY,Jordan:1986ug,CH}. If external fermions
are suppressed, the effective action for our model takes the form,
\begin{eqnarray}
\lefteqn{\Gamma[\phi_{\scriptscriptstyle +},\phi_{\scriptscriptstyle -}] =
S[\phi_{\scriptscriptstyle +}] - S[\phi_{\scriptscriptstyle -}] 
-\frac12 \int \!\! d^4x \! \int \!\! d^4x' } \nonumber \\
& & \times \left\{\matrix{\! 
\phi_{\scriptscriptstyle +}(x) M^2_{\scriptscriptstyle ++}\!(x;x') 
\phi_{\scriptscriptstyle +}(x') + \phi_{\scriptscriptstyle +}(x) 
M^2_{\scriptscriptstyle +-}\!(x;x') \phi_{\scriptscriptstyle -}(x') \! \cr
\!+ \phi_{\scriptscriptstyle -}(x) M^2_{\scriptscriptstyle -+}\!(x;x') 
\phi_{\scriptscriptstyle +}(x') + \phi_{\scriptscriptstyle -}(x) 
M^2_{\scriptscriptstyle --}\!(x;x') \phi_{\scriptscriptstyle -}(x') \!}
\right\} + O(\phi^3_{\pm}) , \qquad
\end{eqnarray}
where $S$ is the free scalar action. The effective field equations are 
obtained by varying with respect to $\phi_+$ and then setting both fields 
equal \cite{CSHY,Jordan:1986ug,CH},
\begin{equation}
\frac{\delta \Gamma[\phi_{\scriptscriptstyle +},\phi_{\scriptscriptstyle -}]
}{\delta \phi_{\scriptscriptstyle +}(x)} \Biggl\vert_{\phi_{\scriptscriptstyle
\pm} = \phi} \!\!\! = \partial_{\mu} \Bigl(a^2 \partial^{\mu} \phi(x)\Bigr) 
- \! \int \! d^4x' \Bigl[M^2_{\scriptscriptstyle ++}\!(x;x') +
M^2_{\scriptscriptstyle +-}\!(x;x')\Bigr] \phi(x') + O(\phi^2) . \label{efe}
\end{equation}
It follows that the two 1PI 2-point functions we need are $M^2_{
\scriptscriptstyle ++}\!(x;x')$ and $M^2_{\scriptscriptstyle +-}\!(x;x')$.
Their sum in (\ref{efe}) gives effective field equations which are causal
in the sense that the two 1PI functions cancel unless $x^{\prime \mu}$ lies
on or within the past light-cone of $x^{\mu}$. Their sum is also real, which
neither 1PI function is separately.

\section{Renormalized One Loop Self-Mass}

In this section we compute and fully renormalize the scalar self-mass-squared 
at one loop order. Our result applies for any scale factor $a(t)$. For the
special case of de Sitter ($a(t) = e^{Ht}$, with constant $H$) we give a
manifestly de Sitter invariant form for $M^2_{\scriptscriptstyle ++}\!(x;x')$ 
and $M^2_{\scriptscriptstyle +-}\!(x;x')$.

Using the Feynman rules of the previous section we see that the $++$ and $+-$ 
contributions from the first diagram in Fig.~2 are,
\begin{equation}
-\Bigl(-if a^D\Bigr) i\Bigl[{}_iS_j\Bigr]_{\scriptscriptstyle +\pm}\!\!\!(x;x') 
\Bigl(\mp i f a^{\prime D} \Bigr) i\Bigl[{}_jS_i\Bigr]_{\scriptscriptstyle
\pm+}\!\!\!(x';x) = \mp \frac{f^2 \Gamma^2(\frac{D}2)}{\pi^D} 
\frac{a a'}{{\Delta x}_{\scriptscriptstyle +\pm}^{2(D-1)}} \; , \label{diag}
\end{equation}
where $a$ is the scale factor at conformal time $\eta$ and $a'$ at $\eta '$.
One cannot yet take the spacetime dimension to $D=4$ because this term is
too singular to give a well-defined integral in (\ref{efe}). To make it
less singular we extract derivatives with respect to the un-integrated
coordinate $x^{\mu}$, which can be pulled outside the integral. The key 
identity can be stated without regard to $\pm$ variations,
\begin{equation}
\frac1{{\Delta x}^{2\alpha}} = \frac1{4(\alpha\!-\!1)(\alpha\!-\!\frac{D}{2})}
\, \partial^2 \Bigl(\frac1{{\Delta x}^{2(\alpha-1)}} \Bigr) \; .
\end{equation}
Two applications of this identity give us,
\begin{equation}
\mp \frac{f^2 \Gamma^2(\frac{D}2)}{\pi^D} \frac{a a'}{{\Delta x}_{
\scriptscriptstyle +\pm}^{2(D-1)}} = \mp
\frac{f^2 a a' \Gamma^2(\frac{D}{2}\!-\!1)}{16 \pi^D (D\!-\!3)(D\!-\!4)} 
\,\partial^4 \Bigl( \frac1{{\Delta x}_{\scriptscriptstyle +\pm}^{2(D-3)}}
\Bigr) \; .
\end{equation}
At this point we could take the limit $D = 4$ were it not for the explicit
factor of $(D\!-\!4)$ in the denominator.

It is now necessary to distinguish the $++$ case, which has a one loop 
counterterm, and $+-$ case, which is free of primitive divergences. 
The trick for obtaining the renormalized result in each case involves 
adding zero using the identities,
\begin{equation}
\partial^2 \Bigl(\frac1{{\Delta x}^{D-2}_{\scriptscriptstyle ++}} \Bigr)
= \frac{i 4 
\pi^{\frac{D}2}}{\Gamma(\frac{D}2 -1)} \delta^D(x-x') \qquad 
{\rm and} \qquad \partial^2 \Bigl(\frac1{{\Delta x}^{D-2}_{\scriptscriptstyle 
+-}} \Bigr)= 0 \; .
\end{equation}
We can therefore write the two self-mass-squared's as,
\begin{eqnarray}
\lefteqn{-iM_{\scriptscriptstyle ++}^2\!(x;x') = -\frac{f^2aa'}{16\pi^D}
\frac{\Gamma^2(\frac{D}{2}\!-\!1)}{(D\!-\!3)(D\!-\!4)} \,\partial^4 \!\left(
\frac{1}{\Delta x_{\scriptscriptstyle ++}^{2(D-3)}}-\frac{\mu^{D-4}}{\Delta
x_{\scriptscriptstyle ++}^{2(\frac{D}{2}-1)}}\right) } \nonumber \\
& & \hspace{.5cm} -\frac{i f^2}{8\pi^{\frac{D}2}} \frac{\Gamma(\frac{D}{2}\!-\!
2)}{(D\!-\!3)}\mu^{D\!-\!4} \, a a' \partial^2 \delta^D(x-x') +
i \delta Z \partial_{\mu} \Bigl(a^{D\!-\!2} \partial^{\mu}\delta^D (x-x')
\Bigr) \qquad \nonumber \\
& & \hspace{.5cm} + i \delta \xi (D\!-\!1) a^{D-2} \Bigl( 2 \frac{\partial^2 
a}{a} + (D-4) \frac{\partial_{\mu} a}{a} \frac{\partial^{\mu} a}{a} \Bigr) 
\delta^D(x-x') + O(f^4) , \qquad \label{prim} \\
\lefteqn{-iM_{\scriptscriptstyle +-}^2\!(x;x') = \frac{f^2aa'}{16\pi^D}
\frac{\Gamma^2(\frac{D}{2}\!-\!1) \, \partial^4}{(D\!-\!3)(D\!-\!4)} \!\left(
\!\frac{1}{\Delta x_{\scriptscriptstyle +-}^{2(D-3)}}-\frac{\mu^{D-4}}{\Delta
x_{\scriptscriptstyle +-}^{2(\frac{D}{2}-1)}} \!\right) \!+\! O(f^4) . \qquad
\qquad \qquad}
\end{eqnarray}
Note the appearance of the dimensional regularization mass scale $\mu$.

By comparing the primitive divergence in (\ref{prim}) with the simple
counterterm (\ref{sum}) that results from the relation (\ref{special}), we
settle on the following choice of counterterms,
\begin{equation}
{\delta \xi} = \frac14 \Bigl(\frac{D-2}{D-1}\Bigr) {\delta Z} + \delta
\xi_{\mbox{\tiny fin}} + O(f^4) \;
{\rm and} \; {\delta Z} = \frac{f^2 \mu^{D-4}}{8 \pi^{\frac{D}2}} 
\frac{\Gamma(\frac{D}2-2)}{(D-3)} + O(f^4) . \quad
\end{equation}
Here $\delta \xi_{\mbox{\tiny fin}}$ is a finite, order $f^2$ contribution 
we shall fix later. With this choice the $++$ result becomes,
\newpage
\begin{eqnarray}
\lefteqn{-iM_{\scriptscriptstyle ++}^2\!(x;x') = -\frac{f^2aa'}{16\pi^D}
\frac{\Gamma^2(\frac{D}{2}\!-\!1)}{(D\!-\!3)(D\!-\!4)} \,\partial^4 \!\left(
\frac{1}{\Delta x_{\scriptscriptstyle ++}^{2(D-3)}}-\frac{\mu^{D-4}}{\Delta
x_{\scriptscriptstyle ++}^{2(\frac{D}{2}-1)}}\right) } \nonumber \\
& & \hspace{3cm} -\frac{i f^2}{8\pi^{\frac{D}2}} \frac{\Gamma(\frac{D}{2}\!-
\!2)}{(D\!-\!3)} \mu^{D\!-\!4} \, \Bigl(a a' - (a a')^{\frac{D}2 - 1} \Bigr) 
\partial^2 \delta^D(x-x') \nonumber \\
& & \hspace{6cm} + i \delta \xi_{\mbox{\tiny fin}} 6 a \partial^2 a \,
\delta^4(x\!-\!x') + O(f^4) \; . \qquad
\end{eqnarray}
Even though the bare scalar is minimally coupled, the divergent parts of
the one loop counterterms are interpretable as the field strength 
renormalization of the conformal kinetic operator! We might have anticipated 
this from the fact that only fermion propagators enter the primitive diagram 
at one loop order, and massless fermions are conformally invariant in any 
dimension. Higher loop diagrams such as those of Fig.~3 involve the scalar 
propagator, which breaks conformal invariance, so we do not expect the 
conformal relation (\ref{special}) between $\delta \xi$ and $\delta Z$ to 
persist at higher loops.

At this stage we take the limit $D=4$ facilitated by the identities,
\begin{eqnarray}
\frac{1}{\Delta x^{2(D-3)}} - \frac{\mu^{D-4}}{\Delta x^{2(\frac{D}{2}-1)}} 
& = & -\Bigl(\frac{D}2 \!-\!2\Bigr) \frac{\ln(\mu^2 \Delta x^2)}{\Delta x^2}
+ O\Bigl( (D\!-\!4)^2\Bigr) \; , \\
a a' - (a a')^{\frac{D}2 - 1} 
& = & -\Bigl(\frac{D}2 \!-\!2\Bigr) a a' \ln(a a') + O\Bigl( (D\!-\!4)^2\Bigr) 
\; .
\end{eqnarray}
The factor of $\ln(a a')$ in the second relation is reminiscent of the
non-local conformal anomaly \cite{DDI} and derives from the same dimensional
mismatch between primitive divergence and counterterm. Putting everything
together and taking the limit $D = 4$ gives,
\begin{eqnarray}
\lefteqn{M^2_{\scriptscriptstyle ++}\!(x;x') = \frac{i f^2 a a'}{32 \pi^4}
\partial^4 \left(\frac{\ln(\mu^2 {\Delta x}^2_{\scriptscriptstyle ++})}{
{\Delta x}^2_{\scriptscriptstyle ++}}\right) } \nonumber \\
& & - \frac{f^2 a a'}{8 \pi^2} \ln(a a') \partial^2 \delta^4(x - x') 
- \delta \xi_{\mbox{\tiny fin}} 6 a \partial^2 a \delta^4(x-x') + O(f^4) \; ,
\label{eq:M++^2} \\
\lefteqn{M^2_{\scriptscriptstyle +-}\!(x;x') = -\frac{i f^2 a a'}{32 \pi^4}
\partial^4 \left(\frac{\ln(\mu^2 {\Delta x}^2_{\scriptscriptstyle +-})}{
{\Delta x}^2_{\scriptscriptstyle +-}}\right) + O(f^4) . } \label{eq:M+-^2}
\end{eqnarray}

We now specialize to de Sitter background, i.e. $a(\eta)=-1/H\eta$
with $H$ constant, to show that the self-mass-squared can be expressed 
in a manifestly de Sitter invariant form. The de Sitter invariant, conformal 
d'Alembertian is,
\begin{equation}
{\cal D}_c \equiv \partial_{\mu} \Bigl(\sqrt{-g} g^{\mu\nu} \partial_{\nu}
\Bigr) - \frac16 \sqrt{-g} R \longrightarrow a \partial^2 a \; .
\label{eq:dSd'Al}
\end{equation}
A simple function of the invariant length $\ell(x;x')$ is,
\begin{equation}
y(x;x') \equiv 4 \sin^2\Bigl(\frac12 H \ell(x;x') \Bigr) = aa' H^2 
\Delta x^2(x;x') \; . \label{eq:dSmodl}
\end{equation}
The polarized forms of this length function are obtained by simply
replacing the coordinate interval, $\Delta x^2(x;x')$, with the interval of
appropriate polarization from (\ref{Dx++^2}) through (\ref{Dx--^2}). Using
these invariants, (\ref{eq:M++^2}) and (\ref{eq:M+-^2}) can be rewritten as 
\begin{eqnarray}
M^2_{\scriptscriptstyle ++}\!(x;x') & = & \frac{i f^2 H^2}{32 \pi^4}
{\cal D}_c {\cal D}_c' \left( \frac{\ln[y_{\scriptscriptstyle ++}(x;x')
\mu^2/H^2]}{y_{\scriptscriptstyle ++}(x;x')}\right) \nonumber \\
& & \hspace{3cm} - \delta \xi_{\mbox{\tiny fin}} R \sqrt{-g} \, \delta^4(x-x') 
+ O(f^4) , \label{dSM++} \\
M^2_{\scriptscriptstyle +-}\!(x;x') & = & - \frac{i f^2 H^2}{32 \pi^4}
{\cal D}_c {\cal D}_c' \left( \frac{\ln[y_{\scriptscriptstyle +-}(x;x')
\mu^2/H^2]}{y_{\scriptscriptstyle +-}(x;x')}\right) + O(f^4) . \qquad
\label{dSM+-}
\end{eqnarray}

\begin{center} 
\begin{picture}(340,100)(0,0)
\DashLine(20,50)(55,50){2}
\Vertex(55,50){2}
\Text(60,50)[l]{$x$}
\DashLine(115,50)(150,50){2}
\Vertex(115,50){2}
\Text(110,51)[r]{$x'$}
\ArrowArc(85,50)(30,0,60)
\ArrowArc(85,50)(30,60,120)
\Vertex(70,76){2}
\Text(65,80)[b]{$y$}
\ArrowArc(85,50)(30,120,180)
\Vertex(100,76){2}
\Text(105,80)[b]{$y'$}
\DashCArc(85,76)(15,0,180){2}
\ArrowArc(85,50)(30,180,360)
\Text(170,46)[b]{\Large $+$}
\DashLine(190,50)(225,50){2}
\Vertex(225,50){2}
\Text(230,50)[l]{$x$}
\DashLine(285,50)(320,50){2}
\Vertex(285,50){2}
\Text(280,51)[r]{$x'$}
\ArrowArc(255,50)(30,0,90)
\Vertex(255,80){2}
\Text(255,85)[b]{$y$}
\ArrowArc(255,50)(30,90,180)
\ArrowArc(255,50)(30,180,270)
\Vertex(255,20){2}
\Text(255,5)[b]{$y'$}
\ArrowArc(255,50)(30,270,360)
\DashLine(255,80)(255,20){2}
\end{picture} 
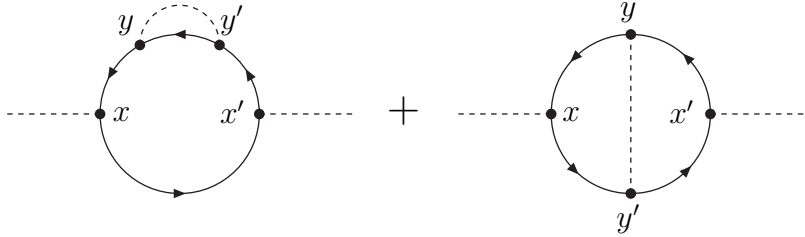
\\ {\rm Fig.~3: Two loop contributions to $M^2(x;x')$.} 
\end{center}

\section{Effective Field Equations}
\label{sec:eft}

In this section we substitute our results (\ref{eq:M++^2}-\ref{eq:M+-^2}) for
$M^2_{\scriptscriptstyle +\pm}\!(x;x')$ into the effective field equation
(\ref{efe}) and work out the result for a spatial plane wave. Most of the
analysis is valid for arbitrary scale factor $a(t)$. Only at the end do
we specialize to the locally de Sitter case and make a choice for $\delta
\xi_{\mbox{\tiny fin}}$ which keeps corrections to the wave functions small 
at one loop order.

The linearized, effective field equation is,
\begin{eqnarray}
\lefteqn{\partial_{\mu} \Bigl(a^2 \partial^{\mu} \phi(x)\Bigr) +
\frac{f^2 a}{8 \pi^2} \left[ \ln(a) \partial^2 \Bigl(a \phi(x)\Bigr) +
\partial^2 \Bigl( \ln(a) a \phi(x)\Bigr) \right] + 
\delta \xi_{\mbox{\tiny fin}} 6 a (\partial^2 a) \phi(x) } \nonumber \\
& &  \hspace{-.7cm} - \frac{i f^2 a}{32 \pi^4} \partial^4 \!\!\int \! d^4x' 
\theta(\eta' \!-\! \eta_I) a' \phi(x') \!\left[ \frac{\ln(\mu^2 {\Delta x}^2_{
\scriptscriptstyle ++})}{{\Delta x}^2_{\scriptscriptstyle ++}} \!-\! 
\frac{\ln(\mu^2 {\Delta x}^2_{\scriptscriptstyle +-})}{{\Delta x}^2_{
\scriptscriptstyle +-}} \right] \!+\! O(f^4) = 0 . \qquad \label{eq:efesub}
\end{eqnarray}
Here $\eta_I$ is the initial conformal time (corresponding to $t=0$) 
at which the state is released in free Bunch-Davies vacuum. The first step
in simplying this equation is to extract another d'Alembertian from the 
nonlocal term in square brackets,
\begin{equation}
\frac{\ln(\mu^2 {\Delta x}^2_{\scriptscriptstyle ++})}{{\Delta x}^2_{
\scriptscriptstyle ++}} - \frac{\ln(\mu^2 {\Delta x}^2_{\scriptscriptstyle +-
})}{{\Delta x}^2_{\scriptscriptstyle +-}} 
= \frac{\partial^2}8 \left[\matrix{\ln^2(\mu^2 {\Delta x}^2_{
\scriptscriptstyle ++}) - 2 \ln(\mu^2 {\Delta x}^2_{\scriptscriptstyle ++}) \cr
- \ln^2(\mu^2 {\Delta x}^2_{\scriptscriptstyle +-}) + 2 \ln(\mu^2 {\Delta x
}^2_{\scriptscriptstyle +-})}\right] \; . \label{nonloc}
\end{equation}
We now define the coordinate intervals $\Delta \eta \equiv \eta \!-\! \eta'$
and $\Delta x \equiv \Vert \vec{x} \!-\! \vec{x}'\Vert$ and recall the $++$
and $+-$ intervals,
\begin{equation}
\Delta x^2_{\scriptscriptstyle ++} = \Delta x^2 - (\vert \Delta \eta\vert
- i \delta)^2 \qquad {\rm and} \qquad \Delta x^2_{\scriptscriptstyle +-} = 
\Delta x^2 - (\Delta \eta + i \delta)^2 \; .
\end{equation}
When $\eta' > \eta$ we have $\Delta x^2_{\scriptscriptstyle ++} =
\Delta x^2_{\scriptscriptstyle +-}$, so the $++$ and $+-$ terms in 
(\ref{nonloc}) cancel. When $\eta' < \eta$ and $\Delta x > \Delta \eta$ 
(spacelike separation) the arguments of the logarithms become positive, real 
numbers for $\delta \rightarrow 0$, so there is also cancellation. Only for 
$\eta' < \eta$ and $\Delta x < \Delta \eta$ (timelike separation) do we 
acquire a nonzero result through the relation,
\begin{equation}
\theta(\Delta \eta - \Delta x) \ln(\mu^2 \Delta x^2_{\scriptscriptstyle + \pm}) 
= \theta(\Delta \eta - \Delta x) \Biggl\{ \ln\Bigl[\mu^2 (\Delta \eta^2 - 
\Delta x^2)\Bigr] \pm i \pi\Biggr\} .
\end{equation}
Hence the square bracketed term in (\ref{eq:efesub}) can be written as,
\begin{equation}
\frac{\ln(\mu^2 {\Delta x}^2_{\scriptscriptstyle ++})}{{\Delta x}^2_{
\scriptscriptstyle ++}} - \frac{\ln(\mu^2 {\Delta x}^2_{\scriptscriptstyle 
+-})}{{\Delta x}^2_{\scriptscriptstyle +-}} 
= \frac{i\pi}2 \partial^2 \Biggl\{ \theta(\Delta \eta -\! \Delta x) 
\Bigl(\ln\Bigl[\mu^2 (\Delta \eta^2 \!-\! \Delta x^2)\Bigr] \!-\! 1\Bigr)\!
\Biggr\} .
\end{equation}
Substituting this relation into (\ref{eq:efesub}) gives the manifestly real
and causal equation,
\begin{eqnarray}
\lefteqn{\partial_{\mu} \Bigl(a^2 \partial^{\mu} \phi(x)\Bigr) +
\frac{f^2 a}{8 \pi^2} \left[ \ln(a) \partial^2 \Bigl(a \phi(x)\Bigr) +
\partial^2 \Bigl( \ln(a) a \phi(x)\Bigr) \right] + 
\delta \xi_{\mbox{\tiny fin}} 6 a (\partial^2 a) \phi(x) } \nonumber \\
& & \hspace{-.5cm} + \frac{f^2 a}{2^6 \pi^3} \partial^6 \!\! \int_{\eta_i}^{
\eta} \!\!\! d\eta' a(\eta') \! \int_{{\Delta x} \leq {\Delta \eta}} 
\!\!\!\!\!\!\!\!\!\!\! d^3x' \phi(\eta',\vec{x}') \Bigl(\ln[\mu^2 ({\Delta 
\eta}^2 \!-\! {\Delta x}^2)] \!-\! 1\Bigr) + O(f^4) = 0 . \qquad \label{randc}
\end{eqnarray}

Because the background geometry is homogeneous, isotropic and spatially flat,
we can build up an arbitrary solution from a superposition of spatial plane
waves of the form,
\begin{equation}
\phi (\eta,\vec{x}) = g(\eta,k) e^{i\vec{k}\cdot\vec{x}} \; . \label{eq:phi}
\end{equation}
Evaluating the derivatives of this in the first two terms of (\ref{randc}) is
straightforward. The nonlocal term, involving the integral, is more 
complicated. To begin we make the change of variable $\vec{y} = \vec{x}' \!-\!
\vec{x}$ and extract the spatial phase factor,
\begin{eqnarray}
\lefteqn{\frac{f^2 a}{2^6 \pi^3} \partial^6 \!\! \int_{\eta_i}^{ \eta} \!\!\! 
d\eta' a(\eta') g(\eta',k) \! \int_{{\Delta x} \leq {\Delta \eta}} 
\!\!\!\!\!\!\!\!\!\!\! d^3x' e^{\vec{k} \cdot \vec{x}'} \Bigl(\ln[\mu^2 
({\Delta \eta}^2 \!-\! {\Delta x}^2)] \!-\! 1\Bigr) } \nonumber \\
& & \hspace{-.7cm} =- \frac{f^2 a}{2^6 \pi^3} e^{i\vec{k} \cdot \vec{x}} 
(\partial_0^2 + k^2)^3 \!\!\! \int_{\eta_i}^{\eta} \!\!\!
d\eta' a(\eta') g(\eta',k) \!\! \int_{\Vert \vec{y} \Vert \leq {\Delta \eta}} 
\!\!\!\!\!\!\!\!\!\!\!\!\!\!\! d^3y \, e^{i \vec{k} \cdot \vec{y}} 
\Bigl(\ln[\mu^2 ({\Delta \eta}^2 \!-\! y^2)] \!-\! 1\Bigr) . \quad
\end{eqnarray}
We next perform the angular integrations and make the change of variables
$y \equiv \Delta \eta z$,
\begin{eqnarray}
\lefteqn{\frac{f^2 a}{2^6 \pi^3} \partial^6 \!\! \int_{\eta_i}^{ \eta} \!\!\! 
d\eta' a(\eta') g(\eta',k) \! \int_{{\Delta x} \leq {\Delta \eta}} 
\!\!\!\!\!\!\!\!\!\!\! d^3x' e^{\vec{k} \cdot \vec{x}'} \Bigl(\ln[\mu^2 
({\Delta \eta}^2 \!-\! {\Delta x}^2)] \!-\! 1\Bigr) } \nonumber \\
& & \hspace{-.5cm} = -\frac{f^2 a}{2^4 \pi^2} e^{i\vec{k} \cdot \vec{x}} 
(\partial_0^2 \!+\! k^2)^3 \!\!\! \int_{\eta_i}^{\eta} \!\!\! d\eta' a(\eta') 
g(\eta',k) \!\! \int_0^{\Delta \eta} \!\!\!\!\! dy \frac{y}{k} \, \sin(ky) 
\ln[\mu^2 ({\Delta \eta}^2 \!-\! y^2)] , \qquad \\
& & \hspace{-.5cm} = -\frac{f^2 a}{2^4 \pi^2} e^{i\vec{k} \cdot \vec{x}} 
(\partial_0^2 \!+\! k^2)^3 \!\!\! \int_{\eta_i}^{\eta} \!\!\! d\eta' a(\eta') 
g(\eta',k) \nonumber \\
& & \hspace{2cm} \times \int_0^1 \! dz \, \Delta \eta^2 \frac{z}{k} \sin(k 
\Delta \eta z)\left[ \ln(\mu^2\Delta \eta^2) + \ln(1 - z^2) -1\right] .
\label{eq:spaceint}
\end{eqnarray}

The integral over $z$ is facilitated by the special function,
\begin{eqnarray}
\xi(\alpha) & \equiv & \int_0^1 dz \, z \sin(\alpha z) \ln(1-z^2) \\
& = & \frac{2}{\alpha^2}\sin\alpha - \frac1{\alpha^2}\left[\cos \alpha
+ \alpha \sin \alpha\right]\left[\mathrm{si}(2\alpha) + \frac{\pi}{2}
\right] \nonumber \\
& & \hspace{1cm} +\frac{1}{\alpha^2}\left[\sin \alpha - \alpha \cos \alpha
\right]
\left[\mathrm{ci}(2\alpha) - \gamma - \ln\left(\frac{\alpha}{2}
\right)\right] \; .
\end{eqnarray}
Here $\gamma$ is the Euler--Mascheroni constant and $\mathrm{si}(x)$ and 
$\mathrm{ci}(x)$ stand for the sine integral and cosine integral
functions,
\begin{eqnarray}
\mathrm{si}(x) & = & - \int_x^{\infty} \!\! dt \, \frac{\sin t}{t} =
-\frac{\pi}{2} + \int_0^x \!\! dt \, \frac{\sin t}{t} \; , \label{si} \\
\mathrm{ci}(x) & = & - \int_x^{\infty} \!\! dt \, \frac{\cos t}{t} =
\gamma + \ln x + \int_0^x \!\! dt \left[ \frac{\cos t - 1 }{t}
\right] \; . \label{ci} 
\end{eqnarray}
Making use of these relations and performing the elementary integrals 
gives,
\begin{eqnarray}
\lefteqn{\frac{f^2 a}{2^6 \pi^3} \partial^6 \!\! \int_{\eta_i}^{ \eta} \!\!\! 
d\eta' a(\eta') g(\eta',k) \! \int_{{\Delta x} \leq {\Delta \eta}} 
\!\!\!\!\!\!\!\!\!\!\! d^3x' e^{\vec{k} \cdot \vec{x}'} \Bigl(\ln[\mu^2 
({\Delta \eta}^2 \!-\! {\Delta x}^2)] \!-\! 1\Bigr) } \nonumber \\
& & \hspace{-.5cm} = -\frac{f^2 a}{2^4 \pi^2} e^{i\vec{k} \cdot \vec{x}} 
\left(\frac{\partial_0^2 \!+\! k^2}{k} \right)^3 \!\!\! \int_{\eta_i}^{\eta} 
\!\!\! d\eta' a(\eta') g(\eta',k) \nonumber \\
& & \hspace{-.2cm} \times \Biggl[\Bigl(\sin(k {\Delta \eta}) \!-\! k {\Delta 
\eta} \cos(k {\Delta \eta}) \Bigr) \Bigl(2 \ln\Bigl(\mu {\Delta \eta} \Bigr) 
\!-\! 1 \Bigr) + (k {\Delta \eta})^2 \xi(k {\Delta \eta}) \Biggr] . \qquad
\label{single}
\end{eqnarray}

The next step is acting the derivatives. Because the integrand in 
(\ref{single}) vanishes at $\eta' = \eta$ like $\ln(\Delta \eta) {\Delta 
\eta}^3$ the first three derivatives commute with the upper limit,
\begin{eqnarray}
\lefteqn{\frac{f^2 a}{2^6 \pi^3} \partial^6 \!\! \int_{\eta_i}^{ \eta} \!\!\! 
d\eta' a(\eta') g(\eta',k) \! \int_{{\Delta x} \leq {\Delta \eta}} 
\!\!\!\!\!\!\!\!\!\!\! d^3x' e^{\vec{k} \cdot \vec{x}'} \Bigl(\ln[\mu^2 
({\Delta \eta}^2 \!-\! {\Delta x}^2)] \!-\! 1\Bigr) } \nonumber \\
& & \hspace{-.5cm} = - \frac{f^2 a}{2^3 \pi^2 k} e^{i \vec{k} \cdot \vec{x}} 
\left(\partial_0^2 + k^2\right) (\partial_0 + i k) (\partial_0 - i k) 
\int_{\eta_i}^{\eta} \!\!\! d\eta' a(\eta') g(\eta',k) \nonumber \\
& & \hspace{-.5cm} \times \!\! \left\{ \sin(k {\Delta \eta}) \!\left[ \ln\!
\left(\frac{2 \mu^2 {\Delta \eta}}{k}\! \right) \!+\! \mathrm{ci}(2 k {\Delta 
\eta}) \!-\! \gamma \right] \!-\! \cos(k {\Delta \eta}) \!\left(\mathrm{si}(2 
k {\Delta \eta}) \!+\! \frac{\pi}{2}\right) \!\! \right\} , \quad \\
& & \hspace{-.5cm} = - \frac{f^2 a}{2^3 \pi^2} e^{i \vec{k} \cdot \vec{x}} 
\left(\partial_0^2 + k^2 \right) (\partial_0 + i k) \int_{\eta_i}^{\eta} 
\!\!\! d\eta' a(\eta') \, g(\eta',k) \, e^{-i k {\Delta \eta}} \nonumber \\
& & \hspace{5cm} \times \left\{ 2 \ln\Bigl(2 \mu {\Delta \eta} \Bigr) +
\int_0^{2 k {\Delta \eta}} \!\!\! dt \left( \frac{e^{it} - 1}{t}\right)
\right\} . \label{eq:58}
\end{eqnarray}
The term containing the logarithm in (\ref{eq:58}) is divergent for $\eta'
= \eta$. We isolate the divergence using,
\begin{equation}
\ln\Bigl(2 \mu {\Delta \eta}\Bigr) = \ln(-2\mu\eta') + \ln\Bigl(1 - 
\frac{\eta}{\eta'}\Bigr) \; .
\end{equation}
We can now act the operator $(\partial_0 \!+\! i k)$ on the non-singular terms
in (\ref{eq:58}),
\begin{eqnarray}
\lefteqn{\frac{f^2 a}{2^6 \pi^3} \partial^6 \!\! \int_{\eta_i}^{ \eta} \!\!\! 
d\eta' a(\eta') g(\eta',k) \! \int_{{\Delta x} \leq {\Delta \eta}} 
\!\!\!\!\!\!\!\!\!\!\! d^3x' e^{\vec{k} \cdot \vec{x}'} \Bigl(\ln[\mu^2 
({\Delta \eta}^2 \!-\! {\Delta x}^2)] \!-\! 1\Bigr) = - \frac{f^2 a}{4 \pi^2} 
e^{i \vec{k} \cdot \vec{x}}} \nonumber \\
& & \hspace{-.5cm} \times \! \left(\partial_0^2 
\!+\! k^2\right) \! \left\{ \! \matrix{a(\eta) g(\eta,k) \ln(-2\mu\eta) + i 
\int_{\eta_i}^{\eta} d\eta' a(\eta') g(\eta',k) \frac{1}{\Delta\eta} \sin(k
{\Delta \eta}) \cr + (\partial_0 + i k)\int_{\eta_i}^{\eta} d\eta' a(\eta') 
g(\eta',k) e^{-i k {\Delta \eta}} \ln\left(1 - \frac{\eta}{\eta'}\right)}
\! \right\} \! . \qquad \label{eq:60}
\end{eqnarray}
Had we instead acted $(\partial_0 \!+\! i k)$ before $(\partial_0 \!-\! i k)$ 
the result would be,
\newpage
\begin{eqnarray}
\lefteqn{\frac{f^2 a}{2^6 \pi^3} \partial^6 \!\! \int_{\eta_i}^{ \eta} \!\!\! 
d\eta' a(\eta') g(\eta',k) \! \int_{{\Delta x} \leq {\Delta \eta}} 
\!\!\!\!\!\!\!\!\!\!\! d^3x' e^{\vec{k} \cdot \vec{x}'} \Bigl(\ln[\mu^2 
({\Delta \eta}^2 \!-\! {\Delta x}^2)] \!-\! 1\Bigr) = - \frac{f^2 a}{4 \pi^2} 
e^{i \vec{k} \cdot \vec{x}}} \nonumber \\
& & \hspace{-.5cm} \times \! \left(\partial_0^2 
\!+\! k^2\right) \! \left\{ \!\matrix{a(\eta) g(\eta,k) \ln(-2\mu\eta) - i 
\int_{\eta_i}^{\eta} d\eta' a(\eta') g(\eta',k) \frac{1}{\Delta\eta} \sin(k
{\Delta \eta}) \cr + (\partial_0 - i k)\int_{\eta_i}^{\eta} d\eta' a(\eta') 
g(\eta',k) e^{i k {\Delta \eta}} \ln\left(1 - \frac{\eta}{\eta'}\right)}
\! \right\} \! . \qquad \label{eq:61}
\end{eqnarray}
A simpler expression results from adding half of (\ref{eq:60}) with 
half of (\ref{eq:61}), 
\begin{eqnarray}
\lefteqn{\frac{f^2 a}{2^6 \pi^3} \partial^6 \!\! \int_{\eta_i}^{ \eta} \!\!\! 
d\eta' a(\eta') g(\eta',k) \! \int_{{\Delta x} \leq {\Delta \eta}} 
\!\!\!\!\!\!\!\!\!\!\! d^3x' e^{\vec{k} \cdot \vec{x}'} \Bigl(\ln[\mu^2 
({\Delta \eta}^2 \!-\! {\Delta x}^2)] \!-\! 1\Bigr) } \nonumber \\
& & \hspace{-.5cm} = - \frac{f^2 a}{4 \pi^2} e^{i \vec{k} \cdot \vec{x}} 
\left(\partial_0^2 \!+\! k^2 \right) \! \left\{ \!\! \matrix{a(\eta) g(\eta,k)
\ln(-2 \mu \eta) \cr
+ \partial_0\int_{\eta_i}^{\eta} d\eta' a(\eta') g(\eta',k) \cos(k {\Delta
\eta}) \ln\left(1 - \frac{\eta}{\eta'}\right) \cr
+ k\int_{\eta_i}^{\eta} d\eta'a(\eta') g(\eta',k) \sin(k {\Delta \eta})
\ln\left(1 - \frac{\eta}{\eta'}\right)} \!\! \right\} . \quad \label{eq:62}
\end{eqnarray}

The top line of (\ref{eq:62}) is comparable to the local one loop terms in
(\ref{randc}). Extracting it is what we have worked so hard to do. The 
remaining, nonlocal terms make only small contributions at late times. 
They can be simplified by first expanding the trigonometric functions using 
the angular addition formulae, then partially integrating to shield the 
logarithmic singularity, and finally bringing another derivative inside,
\begin{eqnarray}
\lefteqn{\partial_0 \! \int_{\eta_i}^{\eta} \!\!\! d\eta'a(\eta') g(\eta',k)
\cos(k\Delta\eta) \ln\Bigl(\frac{\Delta \eta}{-\eta'}\Bigr)
\!+\! k \! \int_{\eta_i}^{\eta} \!\!\! d\eta'a(\eta') g(\eta',k)
\sin(k\Delta\eta) \ln\Bigl(\frac{\Delta \eta}{-\eta'} \Bigr) } \nonumber \\
& & = \cos(k\eta) \partial_0 \!\! \int_{\eta_i}^{\eta} \!\!\! d\eta'a(\eta') 
g(\eta',k) \cos(k\eta') \ln\Bigl(\frac{\Delta \eta}{-\eta'}\Bigr)
\nonumber \\
& & \hspace{3.5cm} + \sin(k\eta) \partial_0 \!\! \int_{\eta_i}^{\eta} \!\!\! 
d\eta' a(\eta') g(\eta',k) \sin(k\eta') \ln\Bigl(\frac{\Delta \eta}{-\eta'}
\Bigr) . \qquad \\
& & = \ln\Bigl(1 \!-\! \frac{\eta}{\eta_i}\Bigr) \frac{\eta_i}{\eta}
a(\eta_i) \cos\Bigl(k (\eta-\eta_i)\Bigr) g(\eta_i,k) \nonumber \\
& & \hspace{3.5cm} + \frac{1}{\eta}\int_{\eta_i}^{\eta} \!\!\! d\eta'
\ln\Bigl(\frac{\Delta \eta}{-\eta'}\Bigr) \, \frac{\partial}{\partial\eta'}
\Biggl\{\!\eta' a(\eta') \cos(k\Delta\eta) g(\eta',k) \!\Biggr\} . \qquad 
\label{eq:63}
\end{eqnarray}

It is now time to combine terms and give the general result for the
linearized effective field equation (\ref{randc}) specialized to a spatial 
plane wave (\ref{eq:phi}). We denote conformal time derivatives with a prime,
\begin{eqnarray}
\partial_{\mu} \Bigl[a^2 \partial^{\mu} \Bigl(g(\eta,k) e^{i \vec{k} \cdot 
\vec{x}}\Bigr)\Bigr] & = & -a^2 e^{i \vec{k} \cdot \vec{x}} \Bigl( g''
+ 2 \frac{a'}{a} g' + k^2 g \Bigr) \; , \\
\frac1{a} \Bigl(\partial_0^2 + k^2 \Bigr) (a g) & = & g'' + 2 \frac{a}{a} g' 
+ k^2 g \; , \\
\Bigl(\partial_0^2 + k^2\Bigr) \Bigl(\ln(-2\mu \eta) a g\Bigr) & = &
\Bigl(\partial_0^2 + k^2\Bigr) (a g) + \frac{2}{\eta} (a g)' - \frac1{\eta^2} 
a g \; .
\end{eqnarray}
Combining everything, and deleting an overall factor of $-a^2 e^{i \vec{k}
\cdot \vec{x}}$, we obtain,
\begin{eqnarray}
\lefteqn{0 = g'' + 2 \frac{a'}{a} g' + k^2 g + \frac{f^2}{4 \pi^2} \ln(-2 \mu
\eta a) \Bigl[ g'' + 2 \frac{a'}{a} g' + \Bigl(\frac{a''}{a} + k^2\Bigr) g
\Bigr]} \nonumber \\
& & \hspace{-.5cm} + \frac{f^2}{4 \pi^2} \Biggl[ \Bigl(\frac{a'}{a} +
\frac2{\eta}\Bigr) g' + \Bigl(\frac{a^{\prime 2}}{2 a^2} + \frac{a''}{2 a} 
+ \frac{2 a'}{\eta a} - \frac1{\eta^2} \Bigr) g\Biggr] + 6 \, {\delta 
\xi_{\mbox{\tiny fin}}} \frac{a''}{a} g \\
& & \hspace{-.5cm} + \frac{f^2}{4 \pi^2 a} \Bigl(\partial_0^2 + k^2\Bigr)
\left\{ \matrix{ \ln\Bigl(1 \!-\! \frac{\eta}{\eta_i}\Bigr) \frac{\eta_i}{\eta}
a(\eta_{i}) \cos\Bigl(k (\eta \!-\! \eta_{i})\Bigr) g(\eta_i,k) \cr
+\int^{\eta}_{\eta_{i}} d\eta' \ln\Bigl(\frac{\Delta \eta}{-\eta'}\Bigr)
\frac{\partial}{\partial \eta'} \Bigl[ \frac{\eta'}{\eta} a(\eta')
\cos(k {\Delta \eta}) g(\eta',k)\Bigr]} \right\} + O(f^4) . \nonumber
\label{eq:gcefe}
\end{eqnarray}

One can better infer asymptotic behaviors if an extra factor of $a^2$ is 
extracted and the equation is converted to physical time $t$,
\begin{eqnarray}
\lefteqn{\ddot{g} + 3 H \dot{g} + \frac{k^2}{a^2} g + \frac{f^2}{4 \pi^2} 
\ln(-2 \mu \eta a) \Bigl[ \ddot{g} + 3 H \dot{g} + \Bigl(2 H^2 + \dot{H} + 
\frac{k^2}{a^2}\Bigr) g \Bigr]} \nonumber \\
& & \hspace{-.5cm} + \frac{f^2}{4 \pi^2} \Biggl[ \Bigl(H + \frac2{\eta a}
\Bigr) \dot{g} + \Bigl(\frac32 H^2 + \frac34 \dot{H} + \frac{2 H}{\eta a} -
\frac1{\eta^2 a^2} \Bigr) g\Biggr] + \delta \xi_{\mbox{\tiny fin}} \Bigl(
12 H^2 + 6 \dot{H}\Bigr) g \nonumber \\
& & \hspace{-.5cm} + \frac{f^2}{4 \pi^2} \Biggl(\frac{\partial^2}{\partial t^2}
\!+\! 3 H \frac{\partial}{\partial t} \!+\! 2 H^2 \!+\! \dot{H} \!+\! 
\frac{k^2}{a^2} \Biggr) \! \left\{ \ln\Bigl(1 \!-\! \frac{\eta}{\eta_i}\Bigr) 
\frac{\eta_i a_i}{\eta a} \cos\Bigl(k (\eta \!-\! \eta_{i})\Bigr) g(\eta_i,k) 
\right. \nonumber \\
& & \left. +\int^t_0 dt' \ln\Bigl(\frac{\Delta \eta}{-\eta'}\Bigr)
\frac{\partial}{\partial t'} \Bigl[ \frac{\eta' a(t')}{\eta a(t)} 
\cos(k {\Delta \eta}) g(\eta',k)\Bigr] \right\} + O(f^4) = 0 . \qquad
\label{teqn}
\end{eqnarray}
It is also useful to recall the slow roll expansion for 
the conformal time,
\begin{equation}
\eta \equiv -\int_t^{\infty} \frac{dt'}{a(t')} = \frac{-1}{H(t) a(t)}
\left\{1 - \frac{\dot{H}(t)}{H^2(t)} + \dots \right\} .
\end{equation}
Hence the combination $\eta a(t)$ is only slowly varying during inflation.
The tree order mode function approaches a constant at late times so we 
will keep one loop corrections small by making finite renormalizations
to cancel any undifferentiated mode functions which are not suppressed by 
slow roll parameters. The best choice for this is,
\begin{equation}
\mu = \frac12 H(t_i) \equiv \frac12 H_i \qquad {\rm and} \qquad \delta 
\xi_{\mbox{\tiny fin}} = \frac{f^2}{32 \pi^2} \; .
\end{equation}
With these choices the effective mode equation becomes,
\begin{eqnarray}
\lefteqn{\ddot{g} + 3 H \dot{g} + \frac{k^2}{a^2} g + \frac{f^2}{4 \pi^2} 
\ln\Bigl(-\frac{\mu}{H_i} \eta a\Bigr) \Bigl[ \ddot{g} + 3 H \dot{g} + 
\Bigl(2 H^2 + \dot{H} + \frac{k^2}{a^2}\Bigr) g \Bigr]} \nonumber \\
& & \hspace{-.5cm} + \frac{f^2}{4 \pi^2} \Biggl[ \Bigl(H + \frac{2}{\eta a}
\Bigr) \dot{g} + \Bigl(3 H^2 + \frac34 \dot{H} + \frac{2 H}{\eta a} -
\frac1{\eta^2 a^2} \Bigr) g\Biggr] \nonumber \\
& & \hspace{-.5cm} + \frac{f^2}{4 \pi^2} \Biggl(\frac{\partial^2}{\partial t^2}
\!+\! 3 H \frac{\partial}{\partial t} \!+\! 2 H^2 \!+\! \dot{H} \!+\! 
\frac{k^2}{a^2} \Biggr) \! \left\{ \ln\Bigl(1 \!-\! \frac{\eta}{\eta_i}\Bigr) 
\frac{\eta_i a_i}{\eta a} \cos\Bigl(k (\eta \!-\! \eta_{i})\Bigr) g(\eta_i,k) 
\right. \nonumber \\
& & \left. +\int^t_0 dt' \ln\Bigl(\frac{\Delta \eta}{-\eta'}\Bigr)
\frac{\partial}{\partial t'} \Bigl[ \frac{\eta' a(t')}{\eta a(t)} 
\cos(k {\Delta \eta}) g(\eta',k)\Bigr] \right\} + O(f^4) = 0 . \qquad
\label{rteqn}
\end{eqnarray}

The next step is to solve the equation perturbatively,
\begin{equation}
g(\eta,k) = g_0(\eta,k) + \frac{f^2}{4\pi^2} \, g_1(\eta,k) + \dots
\end{equation}
Because the tree order mode function obeys,
\begin{equation}
\ddot{g}_0 + 3 H \dot{g}_0 + \frac{k^2}{a^2} \, g_0 = 0 \; , \label{treeeqn}
\end{equation}
the equation for the one loop correction is,
\begin{eqnarray}
\lefteqn{\ddot{g}_1 \!+\! 3 H \dot{g}_1 \!+\! \frac{k^2}{a^2} g_1 \!=\! 
-\ln\Bigl(-\frac{\mu}{H_i} \eta a\Bigr) \Bigl(2 H^2 \!+\! \dot{H}\Bigr) g_0 
\!-\! \Bigl(H \!+\! \frac{2}{\eta a} \Bigr) \dot{g}_0 } \nonumber \\
& & - \Bigl(3 H^2 \!+\! \frac34 \dot{H} \!+\! \frac{2 H}{\eta a} \!-\!
\frac1{\eta^2 a^2} \Bigr) g_0 - \Biggl(\frac{\partial^2}{\partial t^2} \!+\! 
3 H \frac{\partial}{\partial t} \!+\! 2 H^2 \!+\! \dot{H} \!+\! \frac{k^2}{a^2}
\Biggr) \nonumber \\
& & \hspace{3cm} \times \left\{ \matrix{ \ln\Bigl(1 \!-\! \frac{\eta}{\eta_i}
\Bigr) \frac{\eta_i a_i}{\eta a} \cos\Bigl(k (\eta \!-\! \eta_{i})\Bigr) 
g_0(\eta_i,k) \cr
+ \int^t_0 dt' \ln\Bigl(\frac{\Delta \eta}{-\eta'}\Bigr) \frac{\partial}{
\partial t'} \Bigl[ \frac{\eta' a(t')}{\eta a(t)} \cos(k {\Delta \eta}) 
g_0(\eta',k)\Bigr]} \right\} . \qquad
\label{g1eqn}
\end{eqnarray}
Although the solution to (\ref{treeeqn}) has been obtained for a general 
scale factor $a(t)$ \cite{Tsamis:2002qk} the expression is too complicated 
for the integral in (\ref{g1eqn}) to be evaluated in closed form. For 
the special case of de Sitter background ($a(t) = e^{Ht} = -1/H\eta$, with 
$H$ constant) the tree order mode function is,
\begin{equation}
g_0(\eta,k) = \frac{H}{\sqrt{2 k^3}} \left(1 - \frac{ik}{H a}\right)
\exp\left[\frac{ik}{Ha}\right] = \frac{H}{\sqrt{2 k^3}} \Bigl(1 + i k 
\eta\Bigr) e^{-i k \eta} \; .
\end{equation}
Of course many other terms in the effective mode equation (\ref{teqn}) also 
simplify in de Sitter background,
\begin{eqnarray}
\lefteqn{\ddot{g}_1 \!+\! 3 H \dot{g}_1 \!+\! \frac{k^2}{a^2} g_1 = 
H \dot{g}_0 } \nonumber \\
& & \hspace{-.5cm} -\Biggl(\frac{\partial^2}{\partial t^2} \!+\! 3 H 
\frac{\partial}{\partial t} \!+\! 2 H^2 \!+\! \frac{k^2}{a^2} \Biggr) \! 
\left\{ \!\! \matrix{ \ln\Bigl(1 
\!-\! \frac{\eta}{\eta_i} \Bigr) \cos\Bigl(k (\eta \!-\! \eta_{i})\Bigr) 
g_0(\eta_i,k) \cr
+ \int^t_0 dt' \ln\Bigl(\frac{\Delta \eta}{-\eta'}\Bigr) \frac{\partial}{
\partial t'} \Bigl[ \cos(k {\Delta \eta}) g_0(\eta',k)\Bigr]} \!\! \right\} . 
\qquad \label{eq:ds1efe}
\end{eqnarray}

To begin evaluating the nonlocal term we note the differential identity,
\begin{eqnarray}
\lefteqn{\ln\Bigl(\frac{\Delta \eta}{-\eta'}\Bigr) \frac{\partial}{\partial 
\eta'} \Bigl\{ \cos(k {\Delta \eta}) g_0(\eta',k)\Bigr\} = \frac{\partial}{
\partial \eta'} \left\{ \ln\Bigl(\frac{\Delta \eta}{-\eta'}\Bigr) \cos(k 
{\Delta \eta}) g_0(\eta',k) \right\} } \nonumber \\
& & \hspace{6cm} + \Bigl[\frac1{\eta'} + \frac1{\Delta \eta}\Bigr]
\cos(k {\Delta \eta}) g_0(\eta',k) \; . \qquad \label{1stI}
\end{eqnarray}
The $1/\eta'$ term on the right hand side of (\ref{1stI}) can be expressed
as,
\begin{eqnarray}
\lefteqn{\frac1{\eta'} \cos(k {\Delta \eta}) g_0(\eta',k) = \frac{H}{(2 k)^{
\frac32}} \left\{ e^{i k \eta} \Bigl[\frac{e^{-i 2 k \eta'} \!-\! 1}{\eta'}
\Bigr] \right. } \nonumber \\
& & \hspace{2cm} + \left. \frac{\partial}{\partial \eta'} \Bigl[2 \cos(k \eta) 
\ln(-H\eta') -\frac12 e^{ik (\eta - 2 \eta')} + i k \eta' e^{-i k \eta} \Bigr] 
\right\} . \qquad \label{2ndI}
\end{eqnarray}
A similar expression can be obtained for the $1/{\Delta \eta}$ term on
the right hand side of (\ref{1stI}),
\begin{eqnarray}
\lefteqn{\frac1{\Delta \eta} \cos(k {\Delta \eta}) g_0(\eta',k) = 
\frac{H}{(2 k)^{\frac32}} \left\{ (1 \!+\! i k \eta) e^{-i k \eta} 
\Bigl[\frac{e^{i 2 k {\Delta \eta}} \!-\! 1}{\Delta \eta} \Bigr] \right. } 
\nonumber \\
& & \hspace{1cm} + \left. \frac{\partial}{\partial \eta'} \Bigl[-2 (1 \!+\!
i k \eta) e^{-i k \eta} \ln(H \Delta \eta) + \frac12 e^{ik (\eta - 2 \eta')} -
 i k \eta' e^{-i k \eta} \Bigr] \right\} . \qquad \label{3rdI}
\end{eqnarray}
It follows that the bracketed term in (\ref{eq:ds1efe}) is,
\begin{eqnarray}
\lefteqn{\ln\Bigl(1 \!-\! \frac{\eta}{\eta_i} \Bigr) \cos\Bigl(k (\eta \!-\! 
\eta_{i})\Bigr) g_0(\eta_i,k) + \int_{\eta_i}^{\eta} d\eta' \ln\Bigl(
\frac{\Delta \eta}{-\eta'}\Bigr) \frac{\partial}{\partial \eta'} \Bigl[ 
\cos(k {\Delta \eta}) g_0(\eta',k)\Bigr] } \nonumber \\
& & \hspace{1cm} = \ln(1 \!+\! H \eta) g_0(\eta,k) + \ln(-H\eta) 
\Bigl[ \cos(k \eta) g_0(0,k) - g_0(\eta,k)\Bigr] \nonumber \\
& & \hspace{1.5cm} -\frac12 g_0(0,k) e^{i k \eta}\int_{-2 k \eta}^{-2 k \eta_i} 
\!\!\!\!\!\!\!\!\!\! dz \;\; \Bigl[ \frac{e^{iz} \!-\! 1}{z} \Bigr] + \frac12
g_0(\eta,k) \int_0^{2 k {\Delta \eta}_i} \!\!\!\!\!\!\!\!\!\! dz \;\; \Bigl[ 
\frac{e^{iz} - 1}{z} \Bigr] . \qquad \label{nonsource}
\end{eqnarray}
The integrals in this expression could be written in terms of the sine
and cosine integrals (\ref{si}-\ref{ci}) but it is simpler not to do this.

It remains to substitute the nonlocal source term (\ref{nonsource}) into 
the equation (\ref{eq:ds1efe}) for the one loop mode function. Acting
the derivatives gives a complicated expression which considerable effort
brings to the form,
\begin{eqnarray}
\lefteqn{ H^{-2} \ddot{g}_1 + 3 H^{-1} \dot{g}_1 + \frac{k^2 g_1}{H^2 a^2} 
= (g_0^* - g_0) \ln(a) + H^{-1} \dot{g}_0 \left\{ 1 \!-\! \left( 
\frac{e^{2 i k {\Delta \eta}_i} \!+\! 1}{a - 1}\right) \right\} }
\nonumber \\
& & \hspace{1cm} + g_0 \left\{-2 i k \eta - 2 \ln(1 + H\eta) - \frac{ (1 \!-\! 
i k \eta) e^{2 i k {\Delta \eta}_i} \!+\! 1}{a - 1} + \frac{( e^{2 i k 
{\Delta \eta}_i} \!+\! 1}{2 (a-1)^2} \right\} \nonumber \\
& & \hspace{4cm} +\frac12 g_0^* \int_{-2 k \eta}^{-2 k \eta_i} 
\!\!\!\!\!\!\!\!\!\! dz \;\; \Bigl[ \frac{e^{iz} \!-\! 1}{z} \Bigr] - 
\frac12 g_0 \int_0^{2 k {\Delta \eta}_i} \!\!\!\!\!\!\!\!\!\! dz \;\; 
\Bigl[\frac{e^{iz} - 1}{z} \Bigr] . \qquad
\end{eqnarray}
All the mode functions in this expression are evaluated at $\eta$. From
the asymptotic late time expansion for the mode function,
\begin{equation}
g_0(\eta,k) = \frac{H}{\sqrt{2k^3}} \left\{1 + \frac12 \Bigl(\frac{k}{Ha}
\Bigr)^2 + \frac{i}3 \Bigl(\frac{k}{Ha}\Bigr)^3 + O\Bigl(a^{-4}\Bigr)
\right\} ,
\end{equation}
we see that the leading form of the late time source is,
\begin{equation}
H^{-2} \ddot{g}_1 + 3 H^{-1} \dot{g}_1 + \frac{k^2 g_1}{H^2 a^2} =
\frac{H}{\sqrt{2 k^3}} \left\{-\frac{2i}3 \Bigl(\frac{k}{Ha}\Bigr)^3 \ln(a)
+ O\Bigl(a^{-3}\Bigr) \right\} . \label{answer}
\end{equation}
The solution at late times is straightforward,
\begin{equation}
g_1(\eta,k) \longrightarrow \frac{H}{\sqrt{2 k^3}} \left\{ \frac{i}9 
\Bigl(\frac{k}{Ha}\Bigr)^3 \ln^2(a) + O\Bigl(\ln(a) a^{-3}\Bigr) \right\} ,
\end{equation}
except for possible homogeneous terms which can be absorbed into a further
finite field strength renormalization.

\section{Discussion}

We have computed the fully renormalized scalar self-mass-squared at one
loop (\ref{eq:M++^2}-\ref{eq:M+-^2}) for a general scale factor $a(t)$.
For the special case of de Sitter ($a(t) = e^{Ht}$ with constant
$H$) our results can be written in a manifestly de Sitter invariant form 
(\ref{dSM++}-\ref{dSM+-}). Although the computation was simple on account 
of the conformal invariance of Dirac theory, we have not been able to 
locate a prior result in the literature. 

In any case, our real interest lies in the effect the scalar self-mass-squared
has on the late time behavior of the scalar mode functions. For that purpose 
we derived the Schwinger-Keldysh effective field equation (\ref{randc}) at one 
loop order. When specialized to de Sitter background we were able show that
the theory can be renormalized so that there are no significant corrections
at late times.

The reason for the null result is that one loop contributions (Fig.~2) 
involve only the conformally invariant part of the theory: the fermion 
propagator and the Yukawa coupling. As the introduction explained, 
significant quantum effects during inflation require the participation of 
at least one massless particle which is not conformally invariant. The 
first instance of that for the scalar self-mass-squared would come at two 
loop order through the diagrams of Fig.~3. Evaluating them would be a 
formidable undertaking, but perhaps a possible one in view of the fact 
that the two loop contributions to the scalar self-mass-squared have 
recently been obtained for massless, minimally coupled $\phi^4$ in de 
Sitter background \cite{BOW}.

Recall from the similar analysis of the one loop fermion self-energy
\cite{PW2} that the leading one loop correction to the effective field
equations came in with an extra scale factor compared with the classical
terms. It also had a factor of $f^2$, so we expect the one loop correction
to become comparable to the classical mode function at roughly $f^2 a(t)
\sim 1$. Two loop corrections would be down by an extra factor of $f^2$,
but should not contain any more scale factors. So this is one case in 
which it is actually reliable to solve the effective field equations
non-perturbatively with only one loop corrections!

Of course the physics of the result is that the massless and not 
conformally invariant scalar is catalyzing the production of fermions
\cite{PW2}. The process could be throttled if the scalar quickly develops 
a large enough mass. The work done in this paper shows that no such mass
occurs at one loop order. At two loop order we expect the diagrams of
Fig.~3 to induce corrections to the effective field equation with an 
extra factor of $a^2$ relative to the classical terms. However, they 
will also contain an factor of $f^4$, so they would not become important
until $f^4 a^2 \sim 1$. That is the same time at which the production
of fermions is becoming significant. Because at most one fermion can
be produced in each state, we expect the end of inflation to see essentially
all super-horizon fermion modes fully populated. This is a profound 
difference from the usual picture and it seems there must be important
consequences.

Although we obtained the effective mode equation for an arbitrary
scale factor $a(t)$, we have only solved it for the special case of de
Sitter. For more general scale factors there can be stronger asymptotic
contributions at late times. However, because these vanish for de Sitter,
we conclude that any one loop late time correction to the mode function
must be suppressed by at least one factor of a slow roll parameter. The
fact that scalar-catalyzed fermion production goes to completion ought
therefore to be a general feature of slow roll inflation.

\centerline{\bf Acknowledgments}

This work was partially supported by NSF grant PHY-0244714, DOE contract
DE-FG02-97ER\-41029 and by the Institute for Fundamental Theory at the
University of Florida.

\end{document}